\documentclass[11pt]{article}
\usepackage{amsmath}
\usepackage{amssymb}
\usepackage{amsthm}
\usepackage{amsfonts}
\usepackage{graphicx}
\usepackage{textcomp}
\usepackage{color}
\usepackage{hyperref}
\usepackage{tikz-cd}
\numberwithin{equation}{section}

\newcommand {\R}        {{\mathbb{R}}}

\newcommand {\noi}		{{\noindent}}

\begin{document}

\begin{titlepage}

\title{The Variational Field Equations \\of \\ Cosmological Topologically Massive Supergravity}

\author{ Tekin Dereli\footnote{tdereli@ku.edu.tr}, Cem Yeti\c{s}mi\c{s}o\u{g}lu\footnote{cyetismisoglu@ku.edu.tr} \\ {\small Department of Physics, Ko\c{c} University, 34450 Sar{\i}yer, \.{I}stanbul, Turkey }}

\date{19 July 2020}

\maketitle

\begin{abstract}

\noindent We discuss the formulation of  cosmological topologically massive (simple) supergravity theory in three-dimensional
Riemann-Cartan space-times. We use the language of exterior differential forms and the properties of Majorana spinors on 3-dimensional space-times to explicitly demonstrate 
the local supersymmetry of the action density involved. Coupled field equations that are complete in both of their bosonic and fermionic sectors are derived by a
 first order variational principle subject to a torsion-constraint imposed by the method of Lagrange multipliers.
 We compare these field equations with the partial results given in the literature using a second order variational formalism.
\end{abstract}

\vskip 2cm


\thispagestyle{empty}

\end{titlepage}

\maketitle			
\clearpage 

\section{Introduction}
\setcounter{page}{1}

\noindent Three dimensional field theories often provide useful and interesting toy models to work with. In three dimensions,it is possible to couple the Chern-Simons densities, without breaking gauge invariance, to the usual action densities (Yang-Mills, general relativity, etc.\cite{deser-jackiw-templeton1,deser-jackiw-templeton2}), so that field  theories with propagating massive degrees of freedom may be written down. An important set of examples to such theories are provided by the three dimensional massive gravity theories. The variational field equations of these  theories are derived by augmenting the Einstein-Hilbert action (possibly coupled to other terms as well) by a Lorentzian Chern-Simons term. Some examples of massive gravity theories are studied in references \cite{deser-jackiw-templeton1}-\cite{dereli-yetismisoglu2}. The presence of Chern-Simons term helps generate a propagating massive spin-2 field for the otherwise dynamically trivial Einstein's theory \cite{deser-jackiw-tHooft}. In turn these theories are studied to gain insight for the quantization of gravitational theories. \\

\noi In this paper we will be  focusing our attention to  locally supersymmetric generalization of the topologically massive gravity theory \cite{deser-jackiw-templeton1,deser-jackiw-templeton2}. The cosmological topologically massive gravity is determined by an action that contains the standard Einstein-Hilbert term, plus a cosmological constant and the gravitational Chern-Simons term. Despite having third order field equations in metric components, the theory turns out to be  ghost-free and implies causal propagation and has been extensively studied in the literature. It has a viable quantum description for the wrong sign of the Einstein-Hilbert term \cite{li-song-strominger}. It admits the celebrated BTZ black-hole solution \cite{BTZ1, BTZ2} that is asymptotically $AdS_3$. A formulation using a first order constrained variational formalism in Riemann-Cartan spaces 
in the language of exterior differential forms has been described by us before \cite{dereli-yetismisoglu1,dereli-yetismisoglu2,dereli-tucker1}. \\

\noi The supersymmetric generalisation of TMG without or with a cosmological constant has been formulated for the first time in references \cite{deser-kay} and \cite{deser1}, respectively. These theories are obtained by extending the bosonic action of the TMG theory via adding  appropriate fermionic densities. Both of these models are minimally extended supersymmetric theories, that is, the fermionic terms are expressed in terms of only a single gravitino field in the theory.\footnote{For supersymmetric Chern-Simons theories with extended supersymmetry algebras look at \cite{nishino-gates}.} We would like to point out that these theories as they are were described within a second order formalism. That is, the connection coefficients were assumed to be fixed in terms of the dreibein and gravitino fields and not treated as independent variables. Considerable amount of work followed these initial papers using a second order formalism to deal  with a Hamiltonian formulation  and for obtaining the most general supersymmetric solutions. Further references can be found in  \cite{aragone}-\cite{percacci-seszgin}.  \\

\noi The main purpose of the present paper is to formulate the cosmological topologically massive simple supergravity theory 
using a first order torsion-constrained variational principle
on Riemann-Cartan space-times. Riemann-Cartan space-times are geometries with a metric-compatible connection with dynamical torsion in general, in addition to the usual Riemann curvature. These geometries proved to be natural to formulate supergravity theories\footnote{
Simple supergravity theory in (2+1)-dimensions provides a good example for this fact. Its field equations can be given by geometries which have non-vanishing torsion with vanishing total curvature\cite{howe-tucker}.}. Here we will treat the connection fields as independent variables on the same footing as the co-frame and gravitino fields. We are going to constrain the space-time torsion to its familiar expression in simple supergravity theory by using Lagrange multipliers which are  introduced as new independent variables. Since the local supersymmetry transformations of the connection 1-forms are determined with the torsion tensor  fixed from the variation of the Lagrange multipliers themselves, an explicit supersymmetry transformation law for Lagrange multiplier 1-forms is redundant. We will demonstrate  the local supersymmetry of the proposed action and explicitly derive the complete set of variational field equations. \\

\noi The organization of the paper is as follows. In the second section, we provide mathematical preliminaries that set our conventions and fix the notation for the description of the geometry of our three dimensional Riemann-Cartan space-times and the Majorana spinors that we are going to use. In the third section we  briefly discuss firstly the cosmological supergravity theory. Next building up on this example, we discuss the cosmological topologically massive supergravity theory. The proof of the local supersymmetry of the actions are given in explicit detail. Then we consistently derive the variational field equations that haven't been obtained before in their complete form.  
In the final section we finish with a discussion and some concluding remarks.  

\newpage    
\section{Mathematical Preliminaries}
 
\subsection{Riemann-Cartan Space-times}

A 3-dimensional Riemann-Cartan space-time is determined by the triplet \newline $\{M,g,\nabla\}$ where $M$ is a smooth 3-manifold, modeled on a Lorentzian vector space $(\R^{1,2},\eta)$, equipped with a non-degenerate Lorentzian metric $g$ and a linear connection $\nabla$ that is compatible with the metric, $\nabla g=0$. The metric on $M$ is related to the Lorentzian metric of the model space as
\begin{equation}
g(X_a,X_b)=\eta_{ab}=diag(-,+,+). \label{eq1}
\end{equation}
Here $\{X_a\}$ denotes a set of $g$-orthonormal frames on $M$ which are dual to the co-frame 1-forms $\{e^a\}$ by the canonical pairing $e^b(X_a)=\iota_ae^b=\eta_a^b$. 
$\iota_a:\Lambda^p(M)\to\Lambda^{p-1}(M)$\footnote{$\Lambda(M)=\bigoplus_{i=0}^3\Lambda^p(M)$ stands for the exterior algebra over $M$.}  denotes the interior product operation with respect to a frame vector $X_a$. We will use a shorthand $\iota_{ab\dots} = \iota_{X_a} \iota_{X_b} \dots$ that is totally skew-symmetric in its indices. 
The orientation of $M$ is fixed by the volume 3-form  $*1=e^0\wedge e^1 \wedge e^2$ where $*:\Lambda^p(M)\to\Lambda^{3-p}(M)$ is the Hodge duality operator. The linear connection $\nabla$ on $M$ takes values in the Lorentz algebra and may be expressed in terms of a set of totally anti-symmetric connection 1-forms $\{\omega^a_{\ b}\}$ such that
\begin{equation}
\nabla_{X_a}X_b=\omega^c_{\ a}(X_b)X_c. \label{eq2}
\end{equation}   
The torsion 2-forms and the  curvature 2-forms  of the connection are defined through the first and second Cartan structure equations given by
\begin{equation}
de^a+\omega^a_{\ b}\wedge e^b = T^a,  \label{eq6}
\end{equation}
and 
\begin{equation}
d\omega^a_{\ b}+\omega^a_{\ c}\wedge \omega^c_{\ b} =R^a_{\ b},  \label{eq7}
\end{equation}
respectively. We note that the connection 1-forms may be decomposed uniquely into two parts according to
\begin{equation}
\omega^a_{\ b}=\hat{\omega}^a_{\ b}+K^a_{\ b}, \label{eq3}
\end{equation}
where $\{\hat{\omega}^a_{\ b}\}$ denotes the set of torsion-free Levi-Civita connection 1-forms that satisfy
\begin{equation}
de^a+\hat{\omega}^a_{\ b}\wedge e^b = 0, \label{eq4} 
\end{equation}
and $\{K^a_{\ b}\}$ is the set of contortion 1-forms such that
\begin{equation}
K^a_{\ b}\wedge e^b = T^a. \label{eq5}
\end{equation}
The Bianchi identities 
\begin{equation}
DT^a =R^a_{\ b}\wedge e^b\quad , \quad  DR^a_{\ b}=0. \label{eq8}
\end{equation}
are derived as integrability conditions from the Cartan structure equations.
In the formulas above, $d:\Lambda^p(M)\to \Lambda^{p+1}(M)$ and $D:\Lambda^p(M)\to \Lambda^{p+1}(M)$ denote the exterior derivative and covariant exterior derivative with respect to connection 1-forms $\{\omega^a_{\ b}\}$, respectively. 
We also present the contracted version of the first Bianchi identity (\ref{eq8}) because it will be relevant for our calculations  later:
\begin{align}
2 (\iota_{dc}R_{ab}-\iota_{ba}R_{cd})= \iota_{bdc}(DT_a)+\iota_{abd}(DT_c)+\iota_{acd}(DT_b)+\iota_{acb}(DT_d). \label{eq10}
\end{align}
The contracted Bianchi identity (\ref{eq10}) shows us that in the presence of torsion, the two-two symmetry of the components of the Riemann curvature tensor fails in general. 

\medskip

\noindent The Ricci 1-forms $\{Ric_a\}$ and the curvature scalar $R$ are obtained by the following contractions of the curvature 2-forms:
\begin{equation}
Ric_a=\iota_bR^b_{\ a}, \qquad R=\iota^aRic_a. \label{eq11}
\end{equation}
Finally the Einstein 2-forms $\{G_a\}$ are defined in terms of the Ricci 1-forms and the curvature scalar as
\begin{equation}
G_a= G_{ab}*e^b = *\bigg(Ric_a-\frac{1}{2}Re_a\bigg)=-\frac{1}{2}\epsilon_{abc}R^{bc}. \label{eq12}
\end{equation}
The last equality above shows  that the Einstein 2-forms are directly proportional to the curvature 2-forms and vice versa. 
This property holds true only in 3-dimensions. 

\medskip

\noindent Before we move on to describe the spinors that we will work with, we are going to give some identities regarding the exterior algebra that will be helpful in the following sections: 
\begin{align}
\iota_a \xi&= (-1)^p*(e_a \wedge *\xi), \quad \xi\in \Lambda^p(M) \label{eq19} \\
\epsilon^{abc}\epsilon_{klm}&=-\eta^a_k(\eta^b_l\eta^c_m-\eta^c_l\eta^b_m)+\eta^a_l(\eta^b_k\eta^c_m-\eta^b_m\eta^c_k)-\eta^a_m(\eta^b_k\eta^c_l-\eta^c_l\eta^b_k) \label{eq20} \\
e^a\wedge *e_{kl}&=-\eta^a_k*e_l+\eta^a_l*e_k \label{eq21} \\
\epsilon^{abc}*e_{kl}&=-2\eta_{[kl]}^{ab}e^c+2\eta_{[kl]}^{ac}e^b-2\eta_{[kl]}^{bc}e^a \label{eq30}  
\end{align}
where $\epsilon_{abc}=*e_{abc}$ denotes the totally anti-symmetric Levi-Civita symbol with the choice $\epsilon_{012}=1$ and square bracket around some indices means the normalized total anti-symmetrization of those indices.

\subsection{Majorana Spinors}

\noi The spinor fields on $M$ are related to the spinors on the model space $(\R^{1,2},\eta)$ whose Clifford algebra is denoted by $Cl(1,2)$. We are going to use a real realization generated by the Pauli matrices given by 
\begin{equation}
\gamma_0=\begin{pmatrix}
0 & 1\\ 
-1 & 0
\end{pmatrix}, \quad 
\gamma_1=\begin{pmatrix}
0 & 1\\ 
1 & 0
\end{pmatrix}, \quad 
\gamma_2=\begin{pmatrix}
1 & 0\\ 
0 & -1
\end{pmatrix}. \label{eq49}
\end{equation}
These generators  satisfy the Clifford product rule
\begin{equation}
\gamma_a \gamma_b = \eta_{ab}I + \epsilon_{abc} \gamma^c 
\end{equation}
 so that 
\begin{equation}
\{\gamma_a,\gamma_b\}=2\eta_{ab}I, \quad  [\gamma_a,\gamma_b]=2\epsilon_{abc}\gamma^c. \label{eq50}
\end{equation}
The Clifford algebra $Cl(1,2)$ can be spanned by a basis $\{I,\gamma_a,2\sigma_{ab},\gamma_5\}$ where $I$ is the $2\times2$ identity operator, $\{\sigma_{ab}\}$ and $\gamma_5$ are the Lorentz generators and the volume element, given explicitly by
\begin{equation}
\sigma_{ab}=\frac{1}{4} [\gamma_a,\gamma_b]=\frac{1}{2}\epsilon_{abc}\gamma^c,\quad \gamma_5=\gamma_{0} \gamma_{1}\gamma_{2}=I, \label{eq51}
\end{equation}
respectively. The following identities are satisfied by the generators of the Clifford algebra: 
\begin{equation}
2\gamma_a\sigma_{bc}=\eta_{ab}\gamma_c-\eta_{ac}\gamma_b+\epsilon_{abc}I, \label{eq52}
\end{equation}
\begin{equation}
2\sigma_{ab}\gamma_{c}= -\eta_{ac}\gamma_b +\eta_{bc}\gamma_a+\epsilon_{abc}I, \label{eq53}
\end{equation}
\begin{equation}
[\sigma_{ab},\sigma_{cd}]=-\eta_{ac}\sigma_{bd}+\eta_{ad}\sigma_{bc}+\eta_{bc}\sigma_{ad}-\eta_{bd}\sigma_{ac}. \label{eq55}
\end{equation}
Furthermore the following summation identities are also satisfied:
\begin{align}
\gamma^a\gamma_a=3, \quad &\gamma^a\gamma_b\gamma_a=-\gamma_b, \quad \gamma^a\gamma_b\gamma_c\gamma_a=3\eta_{bc}-2\sigma_{bc}, \quad \gamma^a\sigma_{bc}\gamma_a=-\sigma_{bc}, \nonumber\\
&\gamma^a\sigma_{ab}=\gamma_b, \quad 2\sigma^{ab}\sigma_{ab}=-3, \quad 2\sigma^{ab}\gamma_c\sigma_{ab}=\gamma_c. \label{eq56}
\end{align}

\noi Since the rank-2 and rank-3 elements of the Clifford basis are linearly dependent on the rank-0 and rank-1 elements, we use the basis $\{\gamma_A\}=\{I,\gamma_a\}.$

\medskip

\noi The spin group of our model space, $Spin(1,2)\cong SL(2,\R)$ is the double cover of the local Lorentz group $SO(1,2)$.  $Spin(1,2)$ is generated by the elements $\{I, 2\sigma_{ab}\}$ of the even Clifford subalgebra $Cl_0(1,2)$ and the  spinors carry its irreducible representations. In our case the representation space will be $\R^2$, however, components of the spinors should be odd-Grasmann valued. That is, given $\psi=(\psi_1,\psi_2)^T \in \R^2$, both components are nilpotent and they anti-commute:
\begin{equation}
\psi_{1}^2 = 0 = \psi_{2}^2, \quad  \psi_1\psi_2=-\psi_2\psi_1. \label{eq57}
\end{equation}

\noi An adjoint spinor is defined as an element of the dual space of the spinor space. The map between the representation space $\R^2$ and its dual space $(\R^2)^*$ is given by an anti-symmetric operator
\begin{align}
\mathcal{C}:\R^2 &\to (\R^2)^* \nonumber\\
\psi &\mapsto \bar{\psi}=\psi^T\mathcal{C} \label{eq58}
\end{align}
called the charge conjugation operator. In the Majorana realization that we use, 
\begin{equation}
\mathcal{C}=\gamma^0=\begin{pmatrix}
0 & -1\\ 
1 & 0
\end{pmatrix} \label{eq59}
\end{equation}
which satisfies $\mathcal{C}^{-1}=\mathcal{C}^T=-\mathcal{C}$ and $\mathcal{C}^2=-I$. The inverse map acts on the conjugate spinors from the left and defines charge conjugated spinors
\begin{align}
\mathcal{C}^{-1}:(\R^2)^* &\to \R^2 \nonumber\\
\bar{\psi} &\mapsto \psi_{C}=\mathcal{C}(\bar{\psi})^T. \label{eq60}
\end{align}
Note that $\mathcal{C}^{-1}\mathcal{C}=id$ as expected. 
Because we are using real Clifford generators, our spinors are self-charge conjugate, $\psi_C=\psi$. That is to say, all our spinors will be odd-Grassmann valued Majorana (real) spinors.\\

\noi Furthermore we note that under the action of the charge conjugation operator, the Clifford basis elements $\{\gamma_A\}$ are transposed:
\begin{equation}
\mathcal{C}\gamma_A\mathcal{C}^{-1}=-{\gamma_A}^T. \label{eq62}
\end{equation}

\noi Another useful way to think about the charge conjugation matrix is to consider it as a metric on space of the spinors. Using this property, we may pair spinors to obtain objects which have tensorial behavior under local Lorentz transformations. In fact, in 3-dimensions there are only two spinor bi-linears that one may write: 
\begin{equation}
\bar{\psi}\phi\quad , \quad \bar{\psi}\gamma_a\phi. \label{eq63}
\end{equation}
The first bi-linear is a pseudoscalar and the second one is a Lorentz vector.
The other bi-linears $\bar{\psi}\sigma_{ab}\phi$ and $\bar{\psi}\gamma_5\phi$ may be expressed in terms of the above ones.  Any two arbitrary spinors $\psi$ and $\phi$ satisfy the Majorana flip identities, given by
\begin{equation}
\bar{\psi}\phi=\bar{\phi}\psi \quad ,\quad \bar{\psi}\gamma_a\phi=-\bar{\phi}\gamma_a\psi. \label{eq64}
\end{equation}
The complex conjugation operation is an anti-linear anti-involution acting on the Clifford algebra. Consequently, the spinor bi-linears in (\ref{eq64}) are pure imaginary. 
In order to obtain real quantities instead, we have to introduce a factor of complex unit $i$ into these expressions. Therefore 
\begin{equation}
i(\bar{\psi}\phi) \in \R \quad , \quad i(\bar{\psi}\gamma_a\phi) \in \R^{1,2}. \label{eq65}
\end{equation}
Considering the product of three or more spinors, the order of the products may be arranged according to the Fierz rearrangement formula. Suppose $U$ and $V$ are real valued $2\times 2$ matrices  and $\alpha, \beta, \phi, \psi$ are arbitrary Majorana 2-spinors. Then 
\begin{equation}
(\bar{\alpha}U\beta)(\bar{\phi}V\psi)=-\frac{1}{2}\sum_{A=1,a} (\bar{\alpha}U\gamma^AV\psi)(\bar{\phi}\gamma_A\beta).   \label{eq66}
\end{equation}

\noi When considering spinor fields on $M$, we consider sections of the spin bundle on $M$ which takes values in our spinor space and is acted upon by the spin group $SL(2,\R)$ fibrewise. We define the spin covariant exterior derivative operation that acts on a Majorana spinor valued $p$-form section, for instance the gravitino 1-form
over the  Riemann-Cartan space-time as:
\begin{equation}
D\chi= d\chi+\frac{1}{2}\omega^{ab}\sigma_{ab}\wedge \chi \label{eq67}
\end{equation}
where $\{\omega^{ab}\}$ is a set of metric compatible connection 1-forms on $M$ and $\chi$ is a section of Majorana spinor valued $1$-forms. The Ricci's identity takes the following form on the spinor valued differential form fields:
\begin{equation}
D^2\chi=\frac{1}{2}R^{ab}\sigma_{ab}\wedge\chi \label{eq68}
\end{equation}
where $\{R^{ab}\}$ are the curvature 2-forms on $M$.  

\newpage

\section{Supergravity Theories in 3-Dimensions}


\subsection{Cosmological Supergravity}


\noi The action for cosmological supergravity theory \cite{howe-tucker,dereli-deser}  
\begin{equation}
S[e^a,\omega^{ab},\chi ]=\int_M \mathcal{L}_{CSG} \label{eq70}
\end{equation}
is going to be varied in a first order variational formulation with respect to
the co-frames $\{e^a\}$, connection 1-forms $\{\omega^{ab}\}$ and the Majorana spinor valued 1-form gravitino field $\chi$ taken as independent field variables. 
The Lagrangian density 3-form $\mathcal{L}_{CSG}=\mathcal{L}_{SG}+\mathcal{L}_C$ can be decomposed in terms of 
the action densities for the simple supergravity and the cosmological sectors given by
\begin{align}
\mathcal{L}_{SG}&=-\frac{1}{2}R^{ab}\wedge*e_{ab}-\frac{i}{2}\bar{\chi}\wedge D\chi,\label{eq71} 
\end{align}
and
\begin{align}
\mathcal{L}_C&=\Lambda*1-\frac{im}{4}\bar{\chi}\wedge\gamma\wedge\chi,\label{eq72}
\end{align}
respectively. Here we set the gravitational constant $\kappa=1$,  $\Lambda$ is a cosmological constant and $m$ is a mass parameter.  
The fermionic part of the supergravity and cosmological sectors are the kinetic and non-topological mass terms for the Rarita-Schwinger (gravitino) field. The gravitino field 
$\chi$ and its field strength $D\chi$ are Majorana spinor valued 1- and 2-form fields, respectively:
\begin{equation}
\chi=(\iota_a\chi) e^a=\chi_ae^a, \qquad D\chi=\frac{1}{2}(\iota_{ba}D\chi) e^{ab}=(D\chi)_{[ab]}e^{ab}. \label{eq72.5}
\end{equation}
Furthermore we introduced a gamma matrix valued 1-form  $\gamma=\gamma_a e^a$ to write down the mass term for the gravitino field. 
The cosmological constant and the "mass" of the gravitino field shall be related below via $\Lambda=-m^2$ for local supersymmetry.

\medskip

\noi Then the total variation of the action reads (upto a closed form) 
\begin{align}
\dot{\mathcal{L}}&=\dot{e^a}\wedge\bigg\{ -\frac{1}{2}\epsilon_{abc}R^{bc}+i\frac{m}{4}\bar{\chi}\wedge\gamma_a\chi+\Lambda*e_a\bigg\} \nonumber\\
&+{\dot{\omega}}^{ab} \wedge\bigg\{-\frac{1}{2}\epsilon_{abc}\bigg (T^c-\frac{i}{4}	\bar{\chi}\wedge\gamma^c\chi\bigg) \bigg\}+\dot{\bar{\chi}}\wedge\bigg\{-iD\chi-\frac{im}{2}\gamma\wedge\chi\bigg\}. \label{eq73}
\end{align}
\noi We determine from  this expression the coupled field equations of the cosmological supergravity theory:
\begin{align}
&G_a+\Lambda*e_a +i\frac{m}{4}\bar{\chi}\wedge\gamma_a\chi=0, \label{eq84}\\
&D\chi+\frac{m}{2}\gamma\wedge\chi=0, \label{eq85}\\
&T^a=\frac{i}{4}\bar{\chi}\wedge\gamma^a\chi. \label{eq86}
\end{align}

\noi The (infinitesimal) local supersymmetry transformations of the supergravity multiplet  are given as usual by:
\begin{equation}
\dot{e^a}=i\bar{\alpha}\gamma^a\chi, \qquad \dot{\chi}=2D\alpha +m\gamma \alpha \label{eq74}
\end{equation}
where the local supersymmetry parameter $\alpha = \alpha(x)$ is an arbitrary odd-Grassmann valued Majorana spinor.  
In order to determine  the supersymmetry transformation law for the connection field, we look at the variation of the first Cartan structure equation (\ref{eq6}) which yields
\begin{equation}
{\dot{\omega}_{ab}}\wedge e^b =-i D(\bar{\alpha}\gamma_a \chi )  +  \dot{T_a}=\frac{im}{2}\bar{\alpha}\gamma\gamma_a\wedge\chi-i\bar{\alpha}\gamma_aD\chi=:Z_a. \label{eq75}
\end{equation}
The final simplification follows from the field equations  (\ref{eq86}). 
The solution to the system of equations (\ref{eq75}) is obtained  algebraically as
\begin{equation}
2{\dot{\omega}_{ab}}= \iota_aZ_b-\iota_bZ_a-e^c(\iota_{ab}Z_c). \label{eq76}
\end{equation}
that explicitly yields
\begin{align}
{\dot{\omega}_{ab}}&= \frac{i}{2}\bigg(\bar{\alpha}\gamma_a\iota_b(D\chi)-\bar{\alpha}\gamma_b\iota_a(D\chi)+\bar{\alpha}\gamma\iota_{ab}(D\chi)\bigg) \nonumber\\
&-\frac{im}{2}\bigg(\epsilon_{abc}(\bar{\alpha}\gamma^c\chi)+e_a(\bar{\alpha}\chi_b)-e_b(\bar{\alpha}\chi_a)\bigg). \label{eq77}
\end{align}
Although this does not contribute to the transformation of the action density on-shell, we give the transformation law (\ref{eq77}) for the connection 1-forms for completeness. This result will be relevant when we discuss cosmological topologically massive supergravity theory in what follows. 

\medskip

\noi  We now prove the local supersymmetry of cosmological supergravity theory. 
Let us first consider the contributions that are independent of $m$ in the variations of the action density under our local supersymmetry transformations:
\begin{equation}
{\dot{\mathcal{L}}}_{SG}(m=0) =-\frac{i}{2}\epsilon_{abc}(\bar{\alpha}\gamma^a\chi)\wedge R^{bc}-2iD\bar{\alpha}\wedge D\chi. \label{eq78}
\end{equation}
This particular combination may be shown to add up to a closed form by noting that:
\begin{equation}
-\frac{i}{2}\epsilon_{abc}(\bar{\alpha}\gamma^a\chi)\wedge R^{bc}=-i\bar{\alpha}R^{ab}\sigma_{ab}\wedge \chi, \label{eq79}
\end{equation}
and
\begin{equation}
-2iD\bar{\alpha}\wedge D\chi=i\bar{\alpha}R^{ab}\sigma_{ab}\wedge \chi +d(-2i\bar{\alpha}D\chi). \label{eq80}
\end{equation}
The rest of the contributions ($m \neq 0$) are given by
\begin{align}
&i\Lambda \bar{\alpha}*\gamma\wedge\chi+\frac{im^2}{2}\bar{\alpha}\gamma\wedge\gamma\wedge\chi-\frac{m}{4}(\bar{\alpha}\gamma^a\chi)\wedge(\bar{\chi}\wedge\gamma_a \chi) \nonumber\\
&-imD\bar{\alpha}\wedge\gamma\wedge\chi+im\bar{\alpha}\wedge\gamma\wedge D\chi .\label{eq81}
\end{align}
The first two terms cancel each other out when we set $\Lambda=-m^2$  and use the identity $\gamma\wedge\gamma=2*\gamma$.
The last two terms on the other hand can be combined to give 
\begin{equation}
-imD\bar{\alpha}\wedge\gamma\wedge\chi+im\bar{\alpha}\wedge\gamma\wedge D\chi=-d(im\bar{\alpha}\gamma\wedge\chi)-\frac{m}{4}(\bar{\alpha}\gamma^a\chi)\wedge(\bar{\chi}\wedge\gamma_a \chi). \label{eq82}
\end{equation}
When all the above contributions are put together, we are left with a closed form plus a non-linear spinorial expression
\begin{equation}
-\frac{m}{2}(\bar{\alpha}\gamma^a\chi)\wedge(\bar{\chi}\wedge\gamma_a \chi). \label{eq82b}
\end{equation}
It is not difficult to verify that (\ref{eq82b}) vanishes identically by performing a Fierz rearrangement. However, some care is needed for signs during the Fierz 
rearrangements because  we are dealing with spinor valued differential forms. One must first open up an expression in the co-frame basis, apply the Fierz rearrangement formula to the components and then bring back in the basis forms. The final outcome reads
\begin{equation}
-\frac{m}{2}(\bar{\alpha}\gamma^a\chi)\wedge(\bar{\chi}\wedge\gamma_a \chi)=-\frac{m}{2}(\bar{\alpha}\chi)\wedge(\bar{\chi}\wedge\chi)=0. \label{eq83}
\end{equation}
With this result,  the local supersymmetry of the cosmological supergravity action (\ref{eq70}) is proven.

\newpage

\subsection{Cosmological Topologically Massive Supergravity}

\noi  The action functional of this theory is given by 
\begin{equation}
S[e^a,\omega^{ab},\chi, \lambda_a ]=\int_M  \mathcal{L}_{Total}  \label{eq100}
\end{equation}
that will be varied independently with respect to the co-frames $\{e^a\}$, connection 1-forms $\{\omega^{ab}\}$ and the gravitino 1-form $\chi$ as before.
We further introduce below Lagrange multipliers 1-forms $\{\lambda_a\}$ that are also varied as independent variables.
Now our Lagrangian density 3-form decomposes according to,
\begin{equation}
\mathcal{L}_{Total}=\mathcal{L}_{CS}+\mathcal{L}_{SG}+\mathcal{L}_{C} + \mathcal{L}_{Constraint}
\end{equation}
where  we added on to the cosmological supergravity action density (\ref{eq71}) of the previous section, the topological Chern-Simons density 3-form
\begin{align}
\mathcal{L}_{CS}=\frac{1}{\mu}\bigg(\omega^a_{\ b}\wedge d\omega^b_{\ a}+&\frac{2}{3}\omega^a_{\ b}\wedge \omega^b_{\ c}\wedge\omega^c_{\ a}\bigg)-\frac{i}{\mu}\bigg(D\bar{\chi}\wedge*D\chi+*D\bar{\chi}\wedge\gamma\wedge*D\chi\bigg). \label{eq101} 
\end{align}
Here $\mu$ is a new coupling constant. It should be noted that the fermionic part of the Chern-Simons density  (\ref{eq101}) contains derivatives of order 2 of the gravitino field. This is consistent with the fact that third order derivatives of metric components appear in the usual bosonic part of Chern-Simons density.  Alternatively, the fermionic part of the Chern-Simons density 3-form could have been expressed as
$$
-\frac{i}{\mu}\bigg(2D\bar{\chi}\wedge*D\chi-D\bar{\chi}\wedge\gamma*(\gamma\wedge D\chi)\bigg) \label{eq103}
$$
which seems more suitable for a Hamiltonian description. However, as far as the variations of the action are concerned this form is considerably harder to work with and we 
prefer to use our form of the topological action density. 
We furthermore introduced a set of Lagrange multiplier 1-forms $\{\lambda_a\}$  that appear linearly in the constraint Lagrangian density 
\begin{align} 
\mathcal{L}_{Constraint}&=\bigg(T^a-\frac{i}{4}\bar{\chi}\wedge\gamma^a\chi\bigg) \wedge \lambda_a. \label{eq102}
\end{align}
Then  independent variations of the action relative to the Lagrange multipliers impose the constraint that the space-time torsion 2-forms
are given algebraically by (\ref{eq86}) as in the previous section. The remaining variational field equations are to be solved subject to this Lagrangian constraint.  
In Riemannian space-times, in a similar way, one may  introduce a  constraint term of the form $T^a \wedge \lambda_a$ in the action whose variations with respect to the multipliers  set the space-time torsion to zero in a  first order constrained variational formulation of gravitational theories. However, since we are working with a supergravity theory we don't want the torsion to vanish, but be equal to the  quadratic expression given by (\ref{eq86}). 
As we are going to show later on, the origin of the Cotton tensor which involves third derivatives of the metric components in the Einstein field equations is due to this constraint term. 
 The torsion constraint furthermore ensures that the supersymmetry transformation law (\ref{eq77}) of the connection 1-forms  remains as it is in the previous section. 

\medskip

\noi The variation of the total action with respect to the independent field variables turns out to be, modulo a closed form,   
\begin{align}
\dot{\mathcal{L}}_{Total}&=\dot{e^a}\wedge\bigg\{-\frac{1}{2}\epsilon_{abc}R^{bc}-m^2 *e_a  +i\frac{m}{4} \bar{\chi} \wedge \gamma_a \chi 
\nonumber \\&+ \frac{i}{\mu}\bigg(\iota_aD\bar{\chi}\wedge *D\chi-D\bar{\chi}\iota_a*D\chi+*D\bar{\chi}\wedge\gamma_a*D\chi\nonumber\\
&\qquad\qquad-2\iota_a*D\bar{\chi}\gamma\wedge*D\chi+2\iota_aD\bar{\chi}\wedge*(\gamma\wedge*D\chi)\bigg)+D\lambda_a\bigg\} \nonumber\\
&+{\dot{\omega}^{ab}}\wedge\bigg\{-\frac{1}{2}\epsilon_{abc}\bigg(T^c-\frac{i}{4}\bar{\chi}\wedge\gamma^c\chi\bigg)-\frac{2}{\mu}R_{ab}-
\frac{1}{2} \big ( \lambda_a\wedge e_b - \lambda_b \wedge e_a \big ) \nonumber\\
&\qquad\qquad+\frac{i}{2\mu}\epsilon_{abc}\bigg(\bar{\chi}\wedge\gamma^c*D\chi+\bar{\chi}\wedge\gamma^c*(\gamma\wedge*D\chi)\bigg)\bigg\} \nonumber\\
&+\dot{\bar{\chi}}\wedge\bigg\{-iD\chi-i\frac{m}{2} \gamma \wedge \chi -\frac{2i}{\mu}\bigg(D*D\chi+D*(\gamma\wedge*D\chi)\bigg)+\frac{i}{2}\lambda_a\wedge\gamma^a\chi\bigg\}\nonumber\\
&+{\dot{\lambda}_a}\wedge\bigg\{T^a-\frac{i}{4}\bar{\chi}\wedge\gamma^a\chi \bigg\} . \label{eq104}
\end{align}

\medskip

\noi We will first demonstrate the local supersymmetry of the action (\ref{eq100}) under the usual transformations of the co-frame, connection and gravitino fields given by (\ref{eq74}) and (\ref{eq77}). 
The explicit supersymmetry transformations of the Lagrange multiplier 1-forms $\lambda^a$  are not necessary because transformation of the connection is obtained using the torsion field. Therefore the last term in (\ref{eq104}) does not make any contribution to the variations on-shell. 

\medskip

\noi Under local supersymmetry transformations (\ref{eq74}) and (\ref{eq77}), the variation of the action density decomposes as follows: 
\begin{align}
\dot{\mathcal{L}}_{Total}=\dot{\mathcal{L}}_{CS}+\dot{\mathcal{L}}_{SG}+\dot{\mathcal{L}}_{C}+\dot{\mathcal{L}}_{Constraint},  \label{eq105}
\end{align}
where the contribution $\dot{\mathcal{L}}_{SG}+\dot{\mathcal{L}}_{C}$ from the cosmological supergravity sector is already shown to yield a closed form given by (\ref{eq80}). We are left to deal with contributions coming from the topological sector and the constraint term. The contributions from the constraint term can be seen to produce just a closed form by a straightforward computation:
\begin{align}
\dot{\mathcal{L}}_{Constraint}&=i(\bar{\alpha}\gamma^a\chi)\wedge D\lambda_a+iD\bar{\alpha}\wedge \lambda^a\wedge\gamma_a\chi  -i\frac{m}{2} \bar{\alpha} \gamma \gamma^a \wedge \chi \wedge \lambda_a \nonumber\\
&-\frac{i}{2}\bigg(\bar{\alpha}\gamma^a\iota^b(D\chi)-\bar{\alpha}\gamma^b\iota^a(D\chi)+\bar{\alpha}\gamma\iota^{ab}(D\chi)\bigg)\wedge\lambda_a\wedge e_b \nonumber\\
&-\frac{im}{2}\bigg(\epsilon^{abc}(\bar{\alpha}\gamma_c\chi)+e^a(\bar{\alpha}\chi^b)-e^b(\bar{\alpha}\chi^a)\bigg) \wedge e_b \wedge \lambda_a \nonumber \\ 
&=iD(\bar{\alpha}\lambda^a)\wedge\gamma_a\chi-i\bar{\alpha}\lambda^a\wedge\gamma_aD\chi=id(\lambda^a\wedge \bar{\alpha}\gamma_a\chi). \label{eq106}
\end{align}

\noi  We note that the result (\ref{eq106}) does not depend on the explicit form of the Lagrange multiplier 1-forms. We therefore must only check those contributions coming from the topological Chern-Simons density. 
In order to ease the discussion, we are going to deal  separately with terms obtained when $m=0$ and rest of the terms for $\ m\neq 0$.

\bigskip

\noi  {\sl Case: $m=0$} 

\medskip

\noi The supersymmetry transformation of  the Chern-Simons density gives 
\begin{align}
\dot{\mathcal{L}}_{CS}(m=0) &=-\frac{i}{\mu}\bigg(2\bar{\alpha}\gamma^a\iota^b(D\chi)+\bar{\alpha}\gamma\iota^{ab}(D\chi)\bigg)\wedge R_{ab}-\frac{4i}{\mu}D^2\bar{\alpha}\wedge\bigg(*D\chi\nonumber\\
&+*(\gamma\wedge*D\chi)\bigg)-\frac{1}{\mu}(\bar{\alpha}\gamma^a\chi) \wedge\bigg(\iota_aD\bar{\chi}\wedge*D\chi-D\bar{\chi}\wedge\iota_a*D\chi\nonumber\\
&+*D\bar{\chi}\wedge\gamma_a*D\chi-2\iota_a*D\bar{\chi}\gamma\wedge*D\chi+2\iota_aD\bar{\chi}\wedge*(\gamma\wedge*D\chi)\bigg)\nonumber\\
&-\frac{1}{2\mu}\bigg(2\bar{\alpha}\gamma^a\iota^b(D\chi)+\bar{\alpha}\gamma\iota^{ab}(D\chi)\bigg)\wedge\bigg(\epsilon_{abc}\bar{\chi}\gamma^c\wedge*(D\chi+\gamma\wedge*D\chi)\bigg). \label{eq107}
\end{align} 
Showing that this unpromising expression vanishes in fact is laborious but can be done by pursuing the following steps. First we start by manipulating the third and fourth terms in (\ref{eq107}) to cancel the first two terms which are proportional to the curvature 2-forms. Next we perform a Fierz rearrangement once on the remaining terms to bring them into the form of $(\bar{\alpha}\chi)(D\bar{\chi}D\chi)$. Then we will be able to group terms into three distinct generic types that do not mix with each other. Then we finally show that each of these group of terms  vanish on their own. \\

\noi Let us start by manipulating the third term in (\ref{eq107}):
\begin{align}
-\frac{4i}{\mu}D^2\bar{\alpha}&\wedge*D\chi=\frac{i}{\mu}\epsilon_{abc}(\bar{\alpha}\gamma^c*D\chi)\wedge R^{ab}\nonumber\\
&=\frac{i}{2\mu}(\bar{\alpha}\gamma^c\iota_{lk}D\chi)\epsilon_{abc}*e^{kl}\wedge R^{ab} \nonumber\\
&=-\frac{i}{\mu}(\bar{\alpha}\gamma\iota_{ba}D\chi)\wedge R^{ab}+\frac{2i}{\mu}(\bar{\alpha}\gamma^b\iota_{ba}D\chi)R^{ac}\wedge e_c \nonumber\\
&=-\frac{i}{\mu}(\bar{\alpha}\gamma\iota_{ba}D\chi)\wedge R^{ab}-\frac{1}{\mu}(\bar{\alpha}\gamma^b\iota_{ba}D\chi)\wedge(D\bar{\chi}\wedge\gamma^a\chi). \label{eq108}
\end{align}
Above, in the first equality we used the Ricci's identity (\ref{eq68}), in the second line we opened the gravitino field in terms of the co-frame basis as in (\ref{eq72.5}), in the third line we used the co-frame identity (\ref{eq30}) and in the final line we used the first Bianchi identity (\ref{eq8}) together with the torsion expression (\ref{eq86}). We note that the first term in the final equality in (\ref{eq108}) cancels out the second term in (\ref{eq107}). \\

\noi Now we manipulate the fourth term in (\ref{eq107}):
\begin{align}
-\frac{4i}{\mu}D^2\bar{\alpha}&\wedge*(\gamma\wedge*D\chi)=\frac{i}{\mu}\epsilon_{abc}(\bar{\alpha}\gamma^c\gamma^d\iota_dD\chi)\wedge R^{ab} \nonumber\\
&=\frac{i}{\mu}\epsilon_{abc}(\bar{\alpha}\iota^cD\chi)\wedge R^{ab}-\frac{2i}{\mu}(\bar{\alpha}\gamma_b\iota_aD\chi)\wedge R^{ab} \nonumber\\
&=-\frac{i}{\mu}(\bar{\alpha}D\chi)\wedge\epsilon_{abc}\iota^aR^{bc}-\frac{2i}{\mu}(\bar{\alpha}\gamma_b\iota_aD\chi)\wedge R^{ab} \nonumber\\
&=\frac{1}{\mu}(\bar{\alpha}\gamma_b\iota_aD\chi)\wedge*\iota_a(D\bar{\chi}\wedge\gamma^a\chi)-\frac{2i}{\mu}(\bar{\alpha}\gamma_b\iota_aD\chi)\wedge R^{ab}. \label{eq109}
\end{align}
Above, in the first equality we used Ricci's identity (\ref{eq68}), while in the second line we used the Clifford product rule (\ref{eq50}) together with the co-frame identity (\ref{eq20}). In the third line we distributed the interior product operation in the first term and made use of the fact that a 4-form vanishes identically. In the final equality we made use of the contracted Bianchi identity (\ref{eq10}) to show $\epsilon_{abc}\iota^aR^{bc}=2*(\iota_aDT^a)$ and used the torsion expression (\ref{eq86}). We note that the last term in (\ref{eq109}) cancels the first term in the total variation (\ref{eq107}) and we have no terms remaining proportional to curvature 2-forms.\footnote{By looking at these cancellations, we fixed the relative sign between the bosonic and fermionic terms in Chern-Simons action density (\ref{eq101}).} \\

\noi The remaining terms can be brought into a generic form $(\bar{\alpha}\chi)(D\bar{\chi}D\chi)$ by applying the Fierz rearrangement formula (\ref{eq66}). While doing such calculations, we make ample use of the co-frame identities (\ref{eq20})-(\ref{eq30}). Finally bringing everything back together, the relevant piece of the supersymmetry transform of the topological Chern-Simons density will be put into the following form: 
\begin{align}
\dot{\mathcal{L}}_{CS}(m=0) &=\frac{1}{\mu}\bigg\{(\bar{\alpha}\gamma^a\chi)\wedge\bigg[-\frac{1}{4}(\iota_aD\bar{\chi}\wedge*D\chi )-\frac{1}{2}(D\bar{\chi}\iota_a*D\chi )\bigg] \nonumber\\
&-\frac{1}{4}(\bar{\alpha}\gamma\wedge\chi)\wedge(\iota_aD\bar{\chi}\iota^a*D\chi )-\frac{1}{4}*e^{ab}\wedge(\bar{\alpha}\gamma_a\chi)\wedge(\iota^cD\bar{\chi}\iota_{cb}D\chi) \nonumber\\
&+(\bar{\alpha}\chi)\wedge\bigg[D\bar{\chi}*(\gamma\wedge D\chi)-\frac{1}{4}\iota_aD\bar{\chi}\wedge\gamma\iota^a*D\chi\nonumber\\
&+\frac{5}{4}*D\bar{\chi}\wedge*(\gamma\wedge*D\chi)\bigg]+\frac{1}{4}*e^{ab}\wedge(\bar{\alpha}\chi)\wedge(\iota^cD\bar{\chi}\gamma_a\iota_{cb}D\chi ) \nonumber\\
&+(\bar{\alpha}\gamma^a\chi)\wedge\bigg[-\frac{3}{2}*D\bar{\chi}\wedge\gamma_a*D\chi-\frac{3}{2}\iota_aD\bar{\chi}\wedge*(\gamma\wedge*D\chi)\nonumber\\
&+\frac{9}{4}\iota_a*D\bar{\chi}\gamma\wedge*D\chi+\frac{1}{4}\iota^bD\bar{\chi}\wedge\gamma\iota_{ba}D\chi\bigg]\nonumber\\
&+\frac{1}{4}(\bar{\alpha}\gamma\wedge\chi)\wedge \bigg[*D\bar{\chi}*(\gamma\wedge D\chi)+\iota^aD\bar{\chi}\gamma^b\iota_{ab}D\chi\bigg]\nonumber\\
&+\frac{1}{4}*e^{ab}\wedge(\bar{\alpha}\gamma_a\chi)\wedge(\iota_cD\bar{\chi}\gamma_b\iota^c*D\chi )\nonumber\\
&-\frac{1}{2}\epsilon^{abc}(\bar{\alpha}\gamma_a\chi)\wedge(\iota_bD\bar{\chi}\wedge\gamma_c*D\chi )\bigg\}. \label{eq110}
\end{align}
To show that the right hand side  vanishes, we group the terms into three generic types which read as follows:
\begin{align}
1. \qquad &\big(\bar{\alpha}\gamma\chi\big)\big(D\bar{\chi}D\chi \big) \label{eq111} \\
2. \qquad &\big(\bar{\alpha}\chi\big)\big(D\bar{\chi}\gamma D\chi \big)  \label{eq112}\\
3.\qquad &\big(\bar{\alpha}\gamma\chi\big)\big(D\bar{\chi}\gamma D\chi \big) \label{eq113}
\end{align} 
Again these generic types do not mix with each other under Fierz rearrangements and each group of terms vanish on their own. In particular, the terms of the types (\ref{eq111}) and (\ref{eq112}) can be shown to vanish by expanding each term in the co-frame basis and then using co-frame identities, taking $*1$ out of these expressions. However, when applying the same method for terms of the type (\ref{eq113}), some simplifications occur and we obtain the following combination:
\begin{align}
\frac{1}{\mu}&\bigg\{(\bar{\alpha}\gamma^a\chi)\wedge\bigg[3\iota_a*D\bar{\chi}\gamma\wedge*D\chi-\frac{3}{2}*D\bar{\chi}\wedge\gamma_a*D\chi\nonumber\\
&-\frac{3}{2}\iota_aD\bar{\chi}\wedge*(\gamma\wedge*D\chi)\bigg]+\frac{3}{4}*e^{ab}\wedge(\bar{\alpha}\gamma_a\chi)\wedge(\iota_cD\bar{\chi}\gamma_b\iota^c*D\chi )\bigg\}. \label{eq114}
\end{align}
To show that this combination vanishes we move an interior product operation in the first three terms and use the fact that a 4-form field identically vanishes. Then the resulting combination cancels out the fourth term. 
Thus the local supersymmetry of the Chern-Simons density (\ref{eq107}) under the local supersymmetry transformations (\ref{eq74}) and (\ref{eq77}) with $m=0$ is established.  

\medskip

\noi  {\sl Case: $m \neq 0$} 

\medskip




\noi Now we move on to take care of $m \neq 0$ terms coming from topological sector. They explicitly read
\begin{align}
\frac{im}{2}&\bigg[\epsilon_{abc}(\bar{\alpha}\gamma^c\chi)+2e_a(\bar{\alpha}\chi_b)\bigg] \wedge  \bigg\{\frac{2}{\mu}R^{ab}-\frac{i}{2\mu}\epsilon^{abd}\bigg[\bar{\chi}\wedge\gamma_d*D\chi\nonumber\\
+\bar{\chi}&\wedge\gamma_d*(\gamma\wedge*D\chi)\bigg]\bigg\}+\frac{2im}{\mu}D(\bar{\alpha}{\gamma})\wedge*\bigg[D\chi+\gamma\wedge *D\chi\bigg]\nonumber\\
&=-\frac{m}{2\mu}(\bar{\alpha}\gamma^a\chi)\wedge\bigg[(\bar{\chi}\wedge\gamma_a*D\chi)+(\bar{\chi}\wedge\gamma_a\gamma_b\iota^bD\chi) \bigg]\nonumber\\
&+\frac{m}{2\mu}\epsilon_{abc}e^a\wedge(\bar{\alpha}\chi^b)\bigg[(\bar{\chi}\wedge\gamma^c*D\chi)+(\bar{\chi}\wedge\gamma^c\gamma^d\iota_dD\chi)\bigg]\nonumber\\
&-\frac{m}{\mu}(\bar{\alpha}\chi^a)(\bar{\chi}\wedge\gamma_a D\chi)-\frac{m}{2\mu}(\bar{\alpha}\gamma^a*D\chi)\wedge(\bar{\chi}\wedge\gamma_a\chi)\nonumber\\
&-\frac{m}{2\mu}(\bar{\alpha}\gamma^a\gamma^b\iota_bD\chi)\wedge(\bar{\chi}\wedge\gamma_a\chi)+\frac{im}{\mu}\epsilon_{abc}(\bar{\alpha}\gamma^c\chi)\wedge R^{ab}\nonumber\\
&+\frac{2im}{\mu}D\bar{\alpha}\wedge\gamma\wedge*D\chi+\frac{2im}{\mu}D\bar{\alpha}\wedge\gamma\gamma^a\wedge\iota_aD\chi. \label{eq152}
\end{align}
To show that the right hand side of the above equality adds up to a closed form, we will start by manipulating the very last term. Using $\gamma\gamma^a=e^a+\epsilon^{acb}e_b\gamma_c$, we can write
\begin{align}
\frac{2im}{\mu}&D\bar{\alpha}\wedge\gamma\gamma^a\wedge\iota_aD\chi=\frac{4im}{\mu}D\bar{\alpha}\wedge D\chi+\frac{2im}{\mu}(D\bar{\alpha}\gamma_c\iota_{ba}D\chi)\wedge*e^{ac}\wedge e^b \nonumber\\
&=d\bigg(\frac{4im}{\mu}\bar{\alpha}D\chi\bigg)-\frac{4im}{\mu}\bar{\alpha}D^2\chi-\frac{2im}{\mu}D\bar{\alpha}\gamma^b*e^a\iota_{ba}D\chi\nonumber\\
&=d\bigg(\frac{4im}{\mu}\bar{\alpha}D\chi\bigg)-\frac{im}{\mu}\epsilon_{abc}(\bar{\alpha}\gamma^c\chi)\wedge R^{ab}-\frac{2im}{\mu}D\bar{\alpha}\wedge\gamma\wedge*D\chi. \label{eq153}
\end{align}
Above in the second equality we distributed a covariant derivative in the first term and made use of the identity (\ref{eq21}). Then in the third equality we used the Ricci identity and (\ref{eq68}) with the interior product identity (\ref{eq19}). The result (\ref{eq153}) shows that the last three terms in (\ref{eq152}) combine to yield a closed form. The remaining terms of (\ref{eq152}) can be brought into the form $(\bar{\alpha}D\chi)(\bar{\chi}\chi)$ by Fierzing once. The resulting expression can be shown to vanish identically after taking $*1$ out of each term. We have thus proven, with this final observation, the local supersymmetry of  the cosmological topologically massive supergravity  action (\ref{eq100}).

\bigskip

\noi Now we are ready to derive the complete set of field equations of cosmological topologically massive supergravity. 
The field equations are read off from the variations of the total action (\ref{eq104}) and they consist of  the Einstein field equations 
\begin{align}
G_a -m^2 *e_a + &D\lambda_a
=\frac{i}{\mu}\bigg(-\iota_aD\bar{\chi}\wedge*D\chi+D\bar{\chi}(\iota_a*D\chi) -*D\bar{\chi}\wedge\gamma_a*D\chi \nonumber \\ 
&+2(\iota_a*D\bar{\chi})\gamma\wedge*D\chi-2\iota_aD\bar{\chi}\wedge*(\gamma\wedge*D\chi)\bigg) - i\frac{m}{4} \bar{\chi} \wedge \gamma_a \chi \label{eq115}
\end{align}
together with the gravitino field equation
\begin{align}
D\chi&+\frac{m}{2} \gamma \wedge \chi + \frac{2}{\mu}\bigg(D*D\chi+D*(\gamma\wedge*D\chi)\bigg)-\frac{1}{2}\lambda^a \wedge \gamma_a\chi=0, \label{eq116}  
\end{align}
subject to the constraint that the space-time torsion is given by
\begin{align}
&T^a=\frac{i}{4}\bar{\chi}\wedge\gamma^a\chi. \label{eq118} 
\end{align}
We note that the $D\lambda_a$ term in the Einstein field equations (\ref{eq115}) is precisely the term that produces the Cotton tensor in topologically massive gravity theory. 
It is obtained by solving the Lagrange multiplier 1-forms $\lambda_a$ from the connection field equations 
and then substituting the result back in the other field equations. Here we do the same and solve the connection variational field equations below algebraically
for the Lagrange multiplier 1-forms:
\begin{align}
\lambda_a\wedge e_b-&\lambda_b \wedge e_a=2\Sigma_{ab} \label{eq119}  
\end{align}
where
\begin{align}
\Sigma_{ab}&=-\frac{2}{\mu}R_{ab}+\frac{i}{2\mu}\epsilon_{abc} \bar{\chi}\wedge\gamma^c*(D\chi+\gamma\wedge*D\chi). \label{eq120} 
\end{align}
The result turns out to be 
\begin{equation}
\lambda_a=-\frac{4}{\mu}\bigg(Y_a+\frac{i}{4}W_a\bigg) \label{eq121} 
\end{equation}
where 
\begin{align}
Y_a&=Ric_a-\frac{1}{4}Re_a \label{eq122}
\end{align}
are the Schouten curvature 1-forms and
\begin{align}
W_a&=\iota_a*(\bar{\chi}\wedge\gamma)\big (*D\chi+*(\gamma\wedge*D\chi)\big )+\bar{\chi}\bigg(\iota_a*\big (\gamma\wedge*D\chi+\gamma\wedge*(\gamma\wedge*D\chi)\big )\bigg) \nonumber\\
&-\frac{1}{2}*\bigg(\bar{\chi}\wedge\gamma\wedge*(D\chi+\gamma\wedge*D\chi)\bigg)e_a, \label{eq123}
\end{align}
are the corresponding contributions coming from the  fermionic sector, respectively. 
Having obtained the solution (\ref{eq121}) for Lagrange multiplier 1-forms in hand, we can now write down the final form of the variational field equations of cosmological topologically massive gravity. We have the Einstein field equations 
\begin{align}
&G_a - m^2 *e_a +i \frac{m}{4} \bar{\chi} \wedge \gamma_a \chi \nonumber \\ &-\frac{4}{\mu}DY_a-\frac{i}{\mu}\bigg(DW_a-\iota_aD\bar{\chi}\wedge*D\chi+D\bar{\chi}(\iota_a*D\chi)-*D\bar{\chi}\wedge\gamma_a*D\chi  \nonumber \\
&+2(\iota_a*D\bar{\chi})\gamma\wedge*D\chi-2\iota_aD\bar{\chi}\wedge*(\gamma\wedge*D\chi)\bigg)=0, \label{eq124} 
\end{align}
coupled to the gravitino field equation
\begin{align}
D\chi&+\frac{m}{2} \gamma \wedge \chi + \frac{2}{\mu}\bigg(D*D\chi+D*(\gamma\wedge*D\chi)+\bigg(Y^a+\frac{i}{4}W^a\bigg)\wedge\gamma_a\chi\bigg)=0, \label{eq125}
\end{align}
where the torsion 2-forms of space-time are  
\begin{align}
&T^a=\frac{i}{4}\bar{\chi}\wedge\gamma^a\chi. \label{eq118} 
\end{align}
The Schouten curvature 1-forms $Y_a$ and their fermionic counterparts $W_a$  are given by the expressions (\ref{eq122}) and (\ref{eq123}), respectively. \\


\bigskip

\section{Conclusion}

\medskip 

In the present work we formulate the cosmological topologically massive supergravity theory  using a torsion-constrained first order variational formalism 
in the language of exterior differential forms on three dimensional  Riemann-Cartan space-times. In particular, we regard the connection 1-forms as independent 
field variables thus treating them at the same level as local Lorentz co-frames and the gravitino field. 
However, the space-time torsion is constrained algebraically to its standard form by the method of Lagrange multipliers. This is an essential feature of our approach
giving rise to contributions of the Lagrange multiplier fields in the final set of field equations.  
We first prove the invariance of the action under infinitesimal local supersymmetry transformations of the co-frame, connection and the gravitino fields.
This we did in explicit detail.
We also present and simplify the final set of  variational field equations since the field equations in their complete form  had been lacking in the previous literature. 

\medskip

\noi In particular the field equations that come from the connection variations are solved algebraically for the Lagrange multiplier fields. We substitute them into the coupled Einstein and Rarita-Schwinger field equations which arise from the co-frame and gravitino field variations, respectively.
The terms that appear on their right hand sides are  identified as the Cotton 2-forms and their fermionic counterpart, the so-called, Cottino 2-form.  
We note that the variations of the complete Chern-Simons density imply some further non-linear terms besides those coming from  the Lagrange multipliers.   
We wish to make a few remarks concerning these. 
We read off from the final version of the field equations the  Cotton 2-forms
\begin{align}
C_a= DY_a +\frac{i}{4}DW_a  &- \frac{i}{4} \bigg ( \iota_aD\bar{\chi}\wedge*D\chi-D\bar{\chi}(\iota_a*D\chi)+*D\bar{\chi}\wedge\gamma_a*D\chi  \nonumber \\
&-2(\iota_a*D\bar{\chi})\gamma\wedge*D\chi+2\iota_aD\bar{\chi}\wedge*(\gamma\wedge*D\chi) \bigg), \label{eq300} 
\end{align}
and the Cottino 2-form 
\begin{align}
C&= Y^a\wedge\gamma_a\chi +\frac{i}{4}W^a \wedge\gamma_a\chi  +  D*D\chi+D*(\gamma\wedge*D\chi). \label{eq126}
\end{align} 
Let us discuss the Cotton 2-forms first. The Cotton 2-forms that one obtains in the formulation of topologically massive gravity theory are given by only the first term in (\ref{eq300}). The second term governs the higher order contributions in the gravitino field that is due to the fermionic part of the topological action and contains  second, fourth and sixth powers of the gravitino field. This may be observed by separating the connection 1-forms according to (\ref{eq3}) and expanding the covariant derivatives of the gravitino field as:
\begin{equation}
D\chi=\hat{D}\chi+\frac{i}{8}\bigg[(\bar{\chi}^a\gamma^b-\bar{\chi}^b\gamma^a)\chi+\bar{\chi}^a\gamma\chi^b\bigg]\wedge\sigma_{ab}\chi \label{eq301}
\end{equation}
where $\hat{D}$ denotes the covariant exterior derivative operation with respect to Levi-Civita connection and the second term is the contribution of contortion. An important feature of the Cotton 2-forms in the Riemannian case (that is, with no torsion present in the geometry) is that they are traceless. The trace of Cotton 2-forms can be taken by
wedging them with the co-frame from the left as follows:  
\begin{equation}
e^a \wedge \hat{D}\hat{Y}_a = -d(e^a \wedge \hat{Y}_a) = -d (e^{ab}\hat{Y}_{a,b})=0, \label{eq302}
\end{equation}
because the components of the Schouten 1-forms are symmetric. Again, a hat over a quantity means that it is obtained by using the Levi-Civita connection. Of course this does not hold when there is torsion present in the geometry, however, one is tempted to ask whether this property still holds for the full Cotton 2-forms given by (\ref{eq300}),
provided we take our geometry to be Riemannian. Unfortunately even this is not the case. The trace of the modified Cotton 2-forms read:
\begin{align}
e^a\wedge \hat{C_a}&=e^a \wedge \frac{i}{4}\bigg(\hat{D}\hat{W}_a + \iota_a\hat{D}\bar{\chi}\wedge*\hat{D}\chi-\hat{D}\bar{\chi}(\iota_a*\hat{D}\chi)+*\hat{D}\bar{\chi}\wedge\gamma_a*\hat{D}\chi  \nonumber \\
&\qquad \qquad -2(\iota_a*\hat{D}\bar{\chi})\gamma\wedge*\hat{D}\chi+2\iota_a\hat{D}\bar{\chi}\wedge*(\gamma\wedge*\hat{D}\chi)\bigg) \nonumber\\
&= -\frac{i}{4}\bigg( d(e^a \wedge \hat{W}_a) +\hat{D}\bar{\chi}\wedge*\hat{D}\chi+*\hat{D}\bar{\chi}\wedge\gamma\wedge \hat{D}\chi\bigg) \nonumber\\
&=\frac{i}{4}\bigg(2d(\bar{\chi}^a\gamma_a\hat{D}\chi)-2*\hat{D}\bar{\chi}\wedge\gamma\wedge \hat{D}\chi-\bar{\chi}\wedge\hat{D}*(\hat{D}\chi-\gamma \wedge*\hat{D}\chi)\bigg). \label{eq303}
\end{align}
There are two reasons for the modified Cotton 2-forms to be not traceless. The first and main impediment is due to the fact that the components of the contributions $\{W_a=W_{a,b}e^b\}$ are not symmetric unlike the components of Schouten 1-forms. Explicitly they read:
\begin{align}
W_{a,b}&=-\bigg[(\bar{\chi}^k\gamma^l\iota_{lk}D\chi) +\frac{1}{2}\epsilon_{klm}(\bar{\chi}^k\iota^{ml}D\chi)\bigg]\eta_{ab}+(\bar{\chi}^k\gamma_a\iota_{bk}D\chi)+(\bar{\chi}^k\gamma_b\iota_{ak}D\chi) \nonumber\\
&+\bigg[(\bar{\chi}^k\iota_b^{\ l}D\chi)+(\bar{\chi}_b\iota^{lk}D\chi)\bigg]\epsilon_{akl}+(\bar{\chi}_b\gamma^k\iota_{ka}D\chi)+(\bar{\chi}^k\gamma_k\iota_{ab}D\chi). \label{eq304}
\end{align}
In general the variation of a fermionic action with respect to the co-frame field yields an asymmetric tensor. The asymmetry of (\ref{eq304}) is an example to this fact. The second reason is that we included the terms coming from co-frame variation of the fermionic part of Chern-Simons 3-form. The contribution of these terms to trace is the fermionic part of Chern-Simons 3-form itself as can be seen from the second equality in (\ref{eq303}). This contribution is also asymmetric. One final remark that we will make about the modified Cotton 2-forms is that, due to fermionic contributions they also cease to be symmetric and divergence free when working in a Riemannian geometry. We do not write down the divergence and anti-symmetric part of Cotton 2-forms here because their expressions are not very instructive.\\

\noi The Cottino 2-form, unlike its superpartner, is not discussed abundantly in the literature. Only in the references \cite{gibbons-pope-sezgin, becker-bruillard-downes}, the part that is linear in the gravitino field is discussed. It reads in a Riemannian geometry,
\begin{equation}
\hat{C}= {\hat{D}}*{\hat{D}}\chi+{\hat{D}}*(\gamma\wedge*{\hat{D}}\chi)+{\hat{Y}}^a\wedge\gamma_a\chi \label{eq305}
\end{equation} 
and is  linear only when we are working in a Riemannian geometry. Otherwise the full Cottino 2-form (\ref{eq126}) contains terms that are to the first, third and fifth powers in the gravitino field. Similar to the Cotton 2-forms, the higher order contributions are encoded in the term that is proportional to $\{W_a\}$. Again in the Riemannian context, the linear part (\ref{eq305}) of Cottino 2-forms are $\gamma$-traceless. This is the spinorial version of the Cotton tensor being traceless. The $\gamma$-trace operation is given by wedging the Cottino 2-form from the left with the $\gamma$-matrix valued 1-form $\gamma=\gamma_ae^a$ :
\begin{align}
\gamma \wedge \hat{C}&= \gamma \wedge \hat{D}[*\hat{D}\chi+*(\gamma\wedge*\hat{D}\chi)]+\gamma \wedge \hat{Y}^a\wedge\gamma_a \wedge\chi \nonumber\\
&= -\hat{D}[\gamma\wedge\sigma^{ab}\gamma\iota_{ba}\hat{D}\chi]+\epsilon_{abc}e^a \wedge \hat{Y}^b \wedge \gamma^c\chi \nonumber\\
&= 2\hat{D}^2\chi-*\hat{Ric^a} \wedge \gamma_a \chi + \frac{\hat{R}}{2}*\gamma \wedge \chi =0. \label{eq306} 
\end{align}
When showing $\gamma$-tracelessness, in the second equality we used the fact that the connection is torsion-free together with the identity $*\hat{D}\chi+*(\gamma\wedge*\hat{D}\chi)=\sigma^{ab}\gamma\iota_{ba}\hat{D}\chi$ and the fact that Schouten tensor is symmetric. In the final equality we made use of the curvature identity  (\ref{eq12}) and the Ricci identity (\ref{eq68}). The $\gamma$-tracelessness does not hold at the presence of torsion, but we again calculate the $\gamma$-trace of the full Cottino 2-form (\ref{eq126}) when there is no torsion. The final result is
\begin{align}
\gamma \wedge \hat{C}&=\frac{i}{4}\gamma \wedge \hat{W^a} \wedge \gamma_a \chi \nonumber\\
&=\frac{i}{4}\bigg[\frac{1}{2}\epsilon_{abc}\bigg(\gamma_d\iota^{ab}\hat{D}\chi(\bar{\chi}^c\chi_d)+\iota^{ad}\hat{D}\chi(\bar{\chi}_b\gamma_d\chi_c)\bigg)\nonumber\\
&+\frac{1}{2}\gamma^a\iota_{ab}\hat{D}\chi(\bar{\chi}^b\gamma^c\chi_c)+\gamma^a\iota^{bc}\hat{D}\chi\bigg(\bar{\chi}_b\gamma_a\chi_c-\bar{\chi}_a\gamma_b\chi_c\bigg)\bigg]*1. \label{eq307}
\end{align}  
When calculating the $\gamma$-trace we opened the expression in co-frame basis and Fierzed once to bring every term into the form $D\chi(\bar{\chi}\chi)$. The expression (\ref{eq307}) shows that, like in the case of Cotton 2-forms, the contributions due to terms $\{W_a\}$ spoil the $\gamma$-tracelessness of the linear part of the Cottino 2-forms (\ref{eq126}). \\


\bigskip


\noi Finally, we wish to emphasize once again the importance of the torsion constraint (\ref{eq102}) for our first order variational formulation of the cosmological topologically massive supergravity theory. For instance if torsion constraint hasn't been imposed by the method of Lagrange multipliers, then one would have obtained 
by first order variations a completely different set of field equations. In this new theory the space-time torsion would be dynamical rather than being determined algebraically by the gravitino fields. One can see by looking at the variation field equations that  this would be the case:
\begin{align}
R^{ab} &= \Lambda e^{ab}-\frac{i}{\mu}\epsilon^{abc}\bigg(\iota_cD\bar{\chi}\wedge*D\chi-D\bar{\chi}(\iota_c*D\chi)+*D\bar{\chi}\wedge\gamma_c*D\chi\nonumber\\
&-2(\iota_c*D\bar{\chi})\gamma \wedge *D\chi+\iota_cD\bar{\chi}\wedge*(\gamma\wedge *D\chi)\bigg)-i\frac{m}{4}\epsilon^{abc}\bar{\chi}\wedge\gamma_c\chi \\
T^a&=\frac{i}{4}\bar{\chi}\wedge\gamma_a \chi +\frac{2}{\mu}\epsilon^{abc}R_{bc}+\frac{i}{\mu}\bar{\chi}\wedge\gamma\bigg(*D\chi+*(\gamma\wedge*D\chi)\bigg) \label{eq200}\\
&D\chi + \frac{m}{2} \gamma \wedge \chi +\frac{2}{\mu}D\bigg(*D\chi+*(\gamma\wedge*D\chi)\bigg)=0 .
\end{align}
This is a generalized version of our usual torsion expression (\ref{eq118}) with contributions coming from the topological term. Similar to what we have done before, 
it is possible to solve for the contortion 1-forms from (\ref{eq200}). Then using this result, one can further solve for the supersymmetry transformation of connection 1-forms. It is clear that this new transformation for connection 1-forms will have terms proportional to the Chern-Simons coupling constant $\mu$ at different orders. This would be a  completely different theory both at the level of the action and with different supersymmetry transformations of fields. At this point it is not even apparent whether this action will be invariant or not under supersymmetry transformations because of the highly non-linear terms present in the expression above for the torsion. 

\bigskip

\noi As far as we are aware, the full set of field equations (\ref{eq124}),(\ref{eq125}) and (\ref{eq118}) of the cosmological topologically massive supergravity theory has not been given explicitly before in the literature. In the references \cite{deser-kay} and \cite{deser1}, the field equations of the theory are not discussed. In \cite{gibbons-pope-sezgin}, the Einstein field equation is devoid of fermionic contributions and the gravitino field equation only covers the linear part (\ref{eq126}) of Cottino 2-forms. Furthermore all the equations are written in terms of Levi-Civita connection so the higher order contributions in the gravitino field are omitted. In \cite{becker-bruillard-downes}, higher order contributions to the Einstein field equations are not explicitly given and Cottino tensor is again given by the linear expression (\ref{eq126}). The field equations are Taylor expanded to second order and the exact field equations are not discussed. Lastly in the remaining references regarding TMS and CTMS theories, the field equations are not discussed. In our formulation we achieve to express the full set of field equations consistently from a variational principle. By doing so, we are able to find the contributions coming to the Cotton and Cottino 2-forms. \\

\noi The future directions that one may consider is to look further into the fermionic Chern-Simons term and study its properties in connection to the 3D invariants of supermanifolds. This problem is ambitious but quite interesting on its own. One of the other directions is that using a similar first order constrained variational formalism, looking for formulations of other 3D supergravity theories. One important candidate may be the supersymmetric generalisation of Minimally Massive Gravity theory \cite{bergshoeff et al3} which solves the bulk versus boundary clash problem. Another direction may be to consider the models that generalize the CTMS model by extending the action density \cite{adringa et al, bergshoeff-hohm-rosseel, bergshoeff13} or by having extended supersymmetries \cite{deger1, deger2, deger3} and look for new solutions. 

\section{Acknowledgement}
 We dedicate this work to Stanley Deser whose insights were our inspiration. We thank \"{O}zg\"{u}r Sar{\i}o\u{g}lu for useful comments and discussions.


\end{document}